# Friendscope: Exploring In-the-Moment Experience Sharing on Camera Glasses via a Shared Camera

MOLLY JANE NICHOLAS, University of California, Berkeley, USA
BRIAN A. SMITH*, Snap Inc., USA and Columbia University, USA
RAJAN VAISH*, Snap Inc., USA

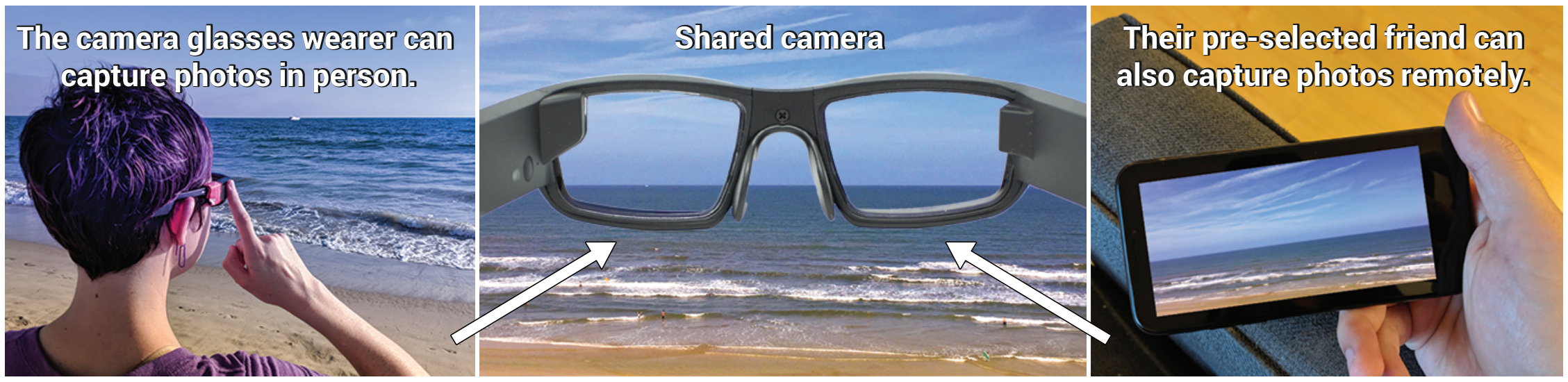

Fig. 1. Overview. Friendscope's key innovation is its *shared camera*, which enables wearers to share control of their camera directly with a pre-selected friend so that both of them can capture photos/videos. The friend is therefore able to receive photos/videos on demand, anytime, as if they are viewing the experience live.

We introduce Friendscope, an instant, in-the-moment experience sharing system for lightweight commercial camera glasses. Friendscope explores a new concept called a *shared camera*. This concept allows a wearer to share control of their camera with a remote friend, making it possible for both people to capture photos/videos from the camera in the moment. Through a user study with 48 participants, we found that users felt connected to each other, describing the shared camera as a more intimate form of livestreaming. Moreover, even privacy-sensitive users were able to retain their sense of privacy and control with the shared camera. Friendscope's different shared camera configurations give wearers ultimate control over who they share the camera with and what photos/videos they share. We conclude with design implications for future experience sharing systems.

CCS Concepts: • **Human-centered computing** → *Ubiquitous and mobile computing systems and tools*; **Social content sharing**; Social media.

Additional Key Words and Phrases: experience sharing, camera glasses, social computing

*Co-Principal Investigators.

## 1 INTRODUCTION

With their light weight, first-person perspective and hands-free form factor, camera glasses have the potential to become a valuable platform for real-time experience sharing. As defined by Bipat et al. [5], camera glasses are glasses with image or video capture functionality only, without the full set of hardware that smartglasses have. Snap Spectacles [66], OhO Waterproof Video Sunglasses [55], and Kestrel Pro [79] are some examples of camera glasses. They look and cost similarly to sunglasses (see Figure 2), and they are one-tenth the weight of full-fledged smartglasses such as Microsoft HoloLens [48] and Magic Leap One [43].

In their recent investigation of camera glasses usage, Bipat et al. [5] found that camera glasses wearers (”Wearers”) use camera glasses for outdoor experiences such as hiking, skiing, and traveling. Notably, Wearers expressed a strong desire to share these experiences with remote family and friends (their ”Friend”), so that they can “see what [the Wearer] is seeing and feeling” and it can be “like they went on the trip with [the Wearer]” [5]. Currently, however, camera glasses primarily support a delayed form of sharing: the Wearer can only share their experience *much later* rather than *in the moment*, and the Friend must simply wait for the Wearer to share highlights from their excursion later. The Wearer needs to (1) record videos, (2) wirelessly transfer them to their phone later, and then (3) send the videos from their phone to the Friend [12, 66, 67]. As a result, both the Wearer's and Friend's sense of togetherness suffers.

In this paper, we introduce *Friendscope*, a system for probing how in-the-moment experience sharing on camera glasses might work. Through Friendscope, we explore a new concept called a *shared camera*, which Figure 1 illustrates. This concept allows the Wearer to share control of their camera directly with a remote Friend. By doing so, Friendscope makes two things possible: the Wearer can capture photos/videos and send them to the Friend instantly, *and* the Friend can capture photos/videos themselves and receive them instantly, on demand, as if they are viewing the experience in the moment. As a result, both the Wearer's and Friend's sense of togetherness can improve.

Friendscope consists of three components that, in tandem, allow the Wearer and their remote Friend to feel connected through the shared camera. Two components are for the Wearer, and one is for their Friend:

- *Camera glasses app (Wearer)*: Allows the Wearer to (a) send photos/videos to their Friend, and (b) receive *trigger requests* (requests to capture photos/videos) from their Friend. The interface does not require a screen, and displays trigger requests using a flashing mock LED. The screenless design minimizes power consumption and makes the app compatible with camera glasses, which do not have screens.
- *Companion smartphone app (Wearer)*: Allows the Wearer to manage the shared camera configuration. Three sharing modes are available: (i) Auto Approve, which automatically approves trigger requests to send photos/videos unless the Wearer declines them; (ii) Manual Approve, where the Wearer must manually approve trigger requests in order to send photos/videos; and (iii) Shared Camera Off, which disables trigger requests entirely.
- *Messaging app (Friend)*: Allows the Friend to send trigger requests to the shared camera and view photos/videos returned from it. Friendscope is compatible with any text messaging app. For this paper, we created our own messaging app.

With Friendscope, we wanted to understand how different shared camera configurations — Auto Approve, Manual Approve, and Shared Camera Off — lead to different types of experiences for users in terms of fellowship, privacy, and sense of control. Shared Camera Off represents our baseline condition, where there is no shared camera at all and only the Wearer can initiate sending photos/videos. This baseline enables us to study the affordances and behaviors that the shared camera and trigger requests bring.

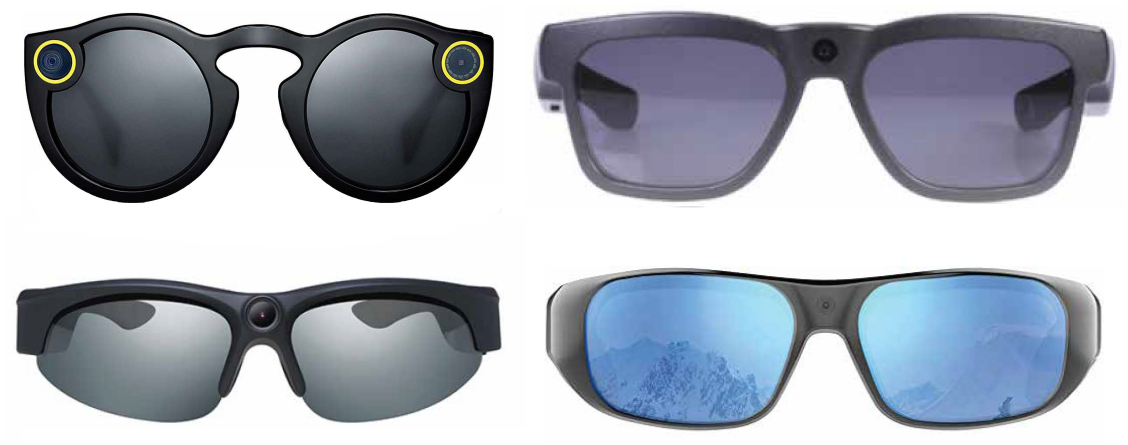

Fig. 2. Examples of camera glasses, Friendscope's target devices. Bipat et al. [5] defined these as lightweight, screenless sunglasses with a very limited set of hardware, typically including a camera, microphone button, and LEDs. These devices are very different from smartglasses, which are more feature-rich and often heavier. Top left: Spectacles [66]. Top right: Kestral Pro [79]. Bottom left: zShades [78]. Bottom right: OhO Waterproof Video Sunglasses [55].

In our main user study with 48 participants, both Wearers and Friends emphasized the sense of being directly connected to each other — in the moment — when the shared camera was enabled (Auto Approve or Manual Approve modes), and described Friendscope as a more intimate form of livestreaming.

Between the three shared camera configurations, Wearers preferred Manual Approve mode the most because it struck the right balance between giving them a sense of fellowship with their Friend and maintaining their own sense of control and privacy over the camera. Because Manual Approve mode provides this balance, participants felt less concerned about privacy violations compared to Auto Approve. Notably, even participants who self-rated as being "highly concerned" with privacy preferred using Friendscope in its Manual Approve mode over Shared Camera Off, the baseline condition representing not having a shared camera at all.

In summary, our main contributions are to enable in-the-moment experience sharing on camera glasses, and to understand the affordances created by different shared camera configurations.

## 2 RELATED WORK

With Friendscope, we explore a new approach to experience sharing which makes in-the-moment experience sharing possible on camera glasses (Figure 2). Here we summarize how experience sharing currently works on camera glasses, as well as how it works on larger devices such as smartglasses and head-mounted displays.

### 2.1 Experience Sharing on Camera Glasses

Camera glasses are a new category of wearable device that look, cost, and weigh similarly to sunglasses (See Figure 2). Their light weight and similarity to sunglasses is precisely why users like them for outdoor activities, "mundane" activities with friends, and while traveling [5]. Consumer product reviews are very positive for glasses with this type of slim form factor [6, 10]. Nearly all of the top uses for camera glasses involve capturing outdoor activities, where network bandwidth is often poor [5]. The success of "action cams" such as GoPro [7, 19], even in spite of their bulky form factors, suggests further that consumers want to share experiences in places with less-than-ideal network bandwidth.

A major limitation of current commercial camera glasses is that they do not support any form of real-time communication. They do not allow a group of people to feel like they are experiencing an event together in the moment. Most of today's consumer camera glasses (including

zShades [78], Kestrel Pro [79], and Spectacles [66]) employ asynchronous sharing to compensate for their constrained hardware, meaning that users share their experiences *later* rather than in the moment. This is because asynchronous sharing is "camera glasses-friendly": it works with camera glasses' form factor and does not require high network connectivity to transmit content well. However, researchers who study live performances have shown that experiences that are shared in the moment increase the sense of connection between participants compared to ones shared later [13, 76].

Haimson and Tang [22] offer an in-depth analysis of what live events on social media platforms must offer in order to be engaging. They show that, to be engaging, live events must offer immediacy (in-the-moment sharing), immersion (first-person perspective), interaction (bi-directional communication), and sociality (fellowship between friends). Camera glasses currently offer immersion and sociality but are missing immediacy and interaction due to their lack of real-time communication between the wearer and their friends. Friendscope addresses these needs by making experience sharing in the moment rather than later and by adding interactivity via the shared camera.

We do not mean to argue that real-time features such as livestreaming or video chatting are impossible to implement on camera glasses. Such features could be implemented, but their technical requirements already strain the hardware of even larger glasses such as Google Glass and would thus be highly impractical for camera glasses. LiKamWa et al. [41] showed that Google Glass can stream video for at most 45 minutes before running out of battery and that doing so will heat the unit by 28° C, risking injury to the wearer. As a result, our work does not explore obvious techniques such as livestreaming to make experience sharing possible on camera glasses. Rather, we explored a new approach — the concept of a shared camera — to make in-the-moment experience sharing possible on camera glasses.

### 2.2 Experience Sharing on Smartglasses and Head-Mounted Displays

In contrast to camera glasses, which are "camera-only" sunglasses with a very limited set of hardware, we use the term smartglasses to refer to more feature-rich and often heavier devices with screens, speakers, and full onboard CPUs. These include Google Glass [18], Vuzix Blade [73, 74], Nreal Light [23], as well as commercial head-mounted displays such as Microsoft HoloLens [48], Oculus Quest [15], and many research prototypes [4, 11, 27, 30, 31, 33, 34, 46, 54, 56, 59, 60].

Most of these devices rely on livestreamed video to enable in-the-moment experience sharing. This is true for both commercial devices and recent research prototypes. The research community has explored many new forms of livestreamed video for experience sharing [4], including 360 degree video [34], blended views [27, 56], and parallel view sharing [33, 56, 60]. But these efforts all employ livestreamed video and require larger devices.

Moreover, researchers have shown that livestreaming approaches do not work well even on high-end devices: they are too hardware-intensive to last long on them. For example, Hashimoto et al. [26] and Paro et al. [57] found, in separate studies, that surgeons rated Google Glass's [18] streaming video quality as poor and inferior to GoPro [57] and iPhone 5 [26], even with their lab environment's high network bandwidth. In addition, LiKamWa et al. [41] showed that streaming video on Google Glass heats the device by a potentially dangerous 28° C and makes the battery last for at most 45 minutes. Both they and Paro et al. [57] conclude that Google Glass is only suitable for notifications and quick snapshots, but not for streaming video.

The main distinction between our work and this existing body of work in experience sharing is that we explore how in-the-moment experience sharing might work on camera glasses. As a result, we propose a very different approach from livestreamed video. In our concept of a shared camera, viewers get continuous access to the Wearer's camera rather than a continuous stream. Our approach is consistent with Neustaedter et al.'s recent findings [54] that viewers value the mere ability to see someone else's experience – the *access* to the experience – and prefer not to watch the entire thing.

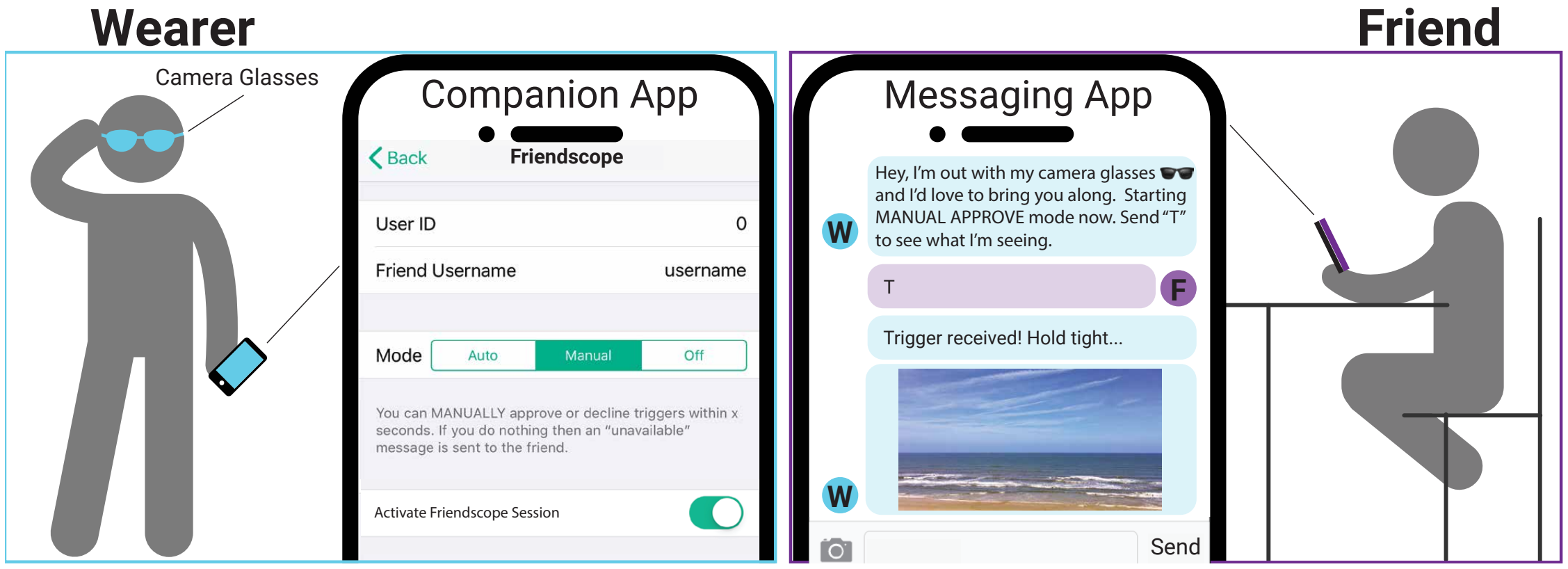

Fig. 3. Friendscope system overview. (Left) The Wearer wears camera glasses and carries an iPhone running the companion app. The companion app allows the user to choose a friend to share with; switch between Auto Approve, Manual Approve, and Shared Camera Off sharing modes; and start/end a Friendscope session. (Right) The remote Friend uses the messaging app to receive photos/videos and to send trigger requests and "thumbs up/down" messages.

### 2.3 Experience Sharing on Other Devices

Conventional videoconferencing systems such as Skype [49], FaceTime [2], and Zoom [80] facilitate co-presence through "talking heads"-style [52] face-to-face communication on mobile, tablet, or laptop devices. Several other systems incorporate wearable cameras and neck- [46], arm- [54, 59], shoulder- [36], or bicycle-mounted [11, 54] smartphones to facilitate experience sharing. Experiences2Go [30] consists of a camcorder and large tablet computer mounted on a tripod. Previous work explores *proxies* or *surrogates* [31, 50, 51, 58], but such surrogate interfaces represent a single-sided approach that provides access to a particular event, compared with our more personalized, bidirectional, experience sharing goals.

## 3 FRIENDSCOPE

Friendscope is an in-the-moment experience sharing system for camera glasses. It employs a novel concept that we call a *shared camera*, which allows the Wearer to share control of their camera with their Friend. With the shared camera, the Wearer can capture *and* send photos/videos to their Friend with a single click, while the Friend can capture *and* receive photos/videos themselves by sending *trigger requests*. Both the Wearer and the Friend can capture photos/videos using the shared camera, and the Friend receives the photos/videos either way. The Friend is able to "join" the Wearer's experience by having continuous access to the shared camera.

When the Friend sends a trigger request, an LED on the inside of the Wearer's camera glasses (which faces the Wearer) flashes green for 10 seconds until it times out. This lets the Wearer know that the Friend would like to receive a photo or video. The Wearer can then approve or decline the trigger request during that period, giving them ultimate control over what is shared. If the Wearer declines a trigger request or the request times out, the Friend receives a message saying the Wearer was unavailable. In both cases, the message is sent at the ten-second timeout moment, affording the Wearer "plausible deniability" [53] that they did not explicitly decline the request. The trigger requests allow Friends to get a photo or video back and see what the Wearer is up to. The Friend can also send the Wearer "thumbs up" and "thumbs down" feedback messages.

We designed Friendscope around the very basic set of hardware components that camera glasses typically have — a camera, a microphone, a button, and an inward-facing LED — so that the system can be applied to a wide range of lightweight, hardware-limited camera glasses and wearable camera interfaces. Our Friendscope implementation has three components: a camera glasses app and a companion smartphone app for the Wearer, and a smartphone messaging app for the Friend. We describe these in detail in the rest of this section.

## 3.1 Wearer Side: Companion Smartphone App

As with most consumer camera glasses [66, 78, 79], Friendscope's settings are controlled by a companion app on the Wearer's smartphone, in our case an iOS app. The app is depicted on the left side of Figure 3. It allows the Wearer to start and end Friendscope sessions and to switch between sharing modes based on their current scenario or comfort level.

*3.1.1 Friendscope sessions.* Camera glasses wearers typically use their glasses in *sessions* while they are performing an activity [5], so we designed Friendscope to be session-based. Using the companion app, the Wearer can start and end Friendscope sessions and specify who to include as their Friend. Pre-selecting the Friend before starting the session makes Friendscope nearly hands-free for the Wearer and gives them complete control over who has access to the shared camera.

When the Wearer starts a session, their Friend receives the following invitation message as shown at the top-right of Figure 3: *"Hey, I'm out with my camera glasses, and I'd love to bring you along. Send 'T' to trigger a photo or video of what I'm seeing right now."* The Friend also receives a message when the Wearer ends a session or changes mode. Together, these messages let the Friend know about the Wearer's availability and whether it is a good time for the Friend to send a trigger request. While a session is active, the Wearer can spontaneously initiate and send photos/videos to show the Friend what they are up to. The Wearer can send photos/videos whenever they want and do not need to receive a trigger request first.

*3.1.2 Sharing modes.* Friendscope's shared camera lets Friends "join" the Wearer's experience, but it can also affect the Wearer's sense of privacy and control over what is captured. As a result, we implemented three *sharing modes* (shared camera configurations) within Friendscope to allow the Wearer to control what happens when the Friend triggers the shared camera. These modes give the Wearer more agency and allow us to study how Wearers manage the tradeoff between having complete control over their camera and giving their Friend easy access to it. The three configurations are:

- `SHARED CAMERA OFF MODE`: This mode disables the shared camera and the Friend's ability to send trigger requests. Shared Camera Off represents our baseline condition where there is no shared camera at all. That is, Friends cannot send trigger requests in this condition, and the Wearer must initiate all photos/videos. Of Friendscope's three sharing modes, this mode gives Wearers the most control but gives the Friend the least amount of access to the Wearer's experience.
- `AUTO APPROVE MODE`: This mode automatically approves trigger requests when they time out, sending the Friend a ten-second video. Since this mode is hands-free, we had to choose either photos or videos to be automatically sent. During our pilot study with 22 participants, we found that Friends preferred videos to be sent as the default in this mode. Before timeout, the Wearer can either decline the trigger request or fulfill the trigger request "early" by capturing a photo/video. This mode allows the Wearer to fulfill trigger requests hands-free but gives them the least amount of control compared to the other modes. The Friend, on the other hand, gets direct access to the Wearer's experience with this mode.

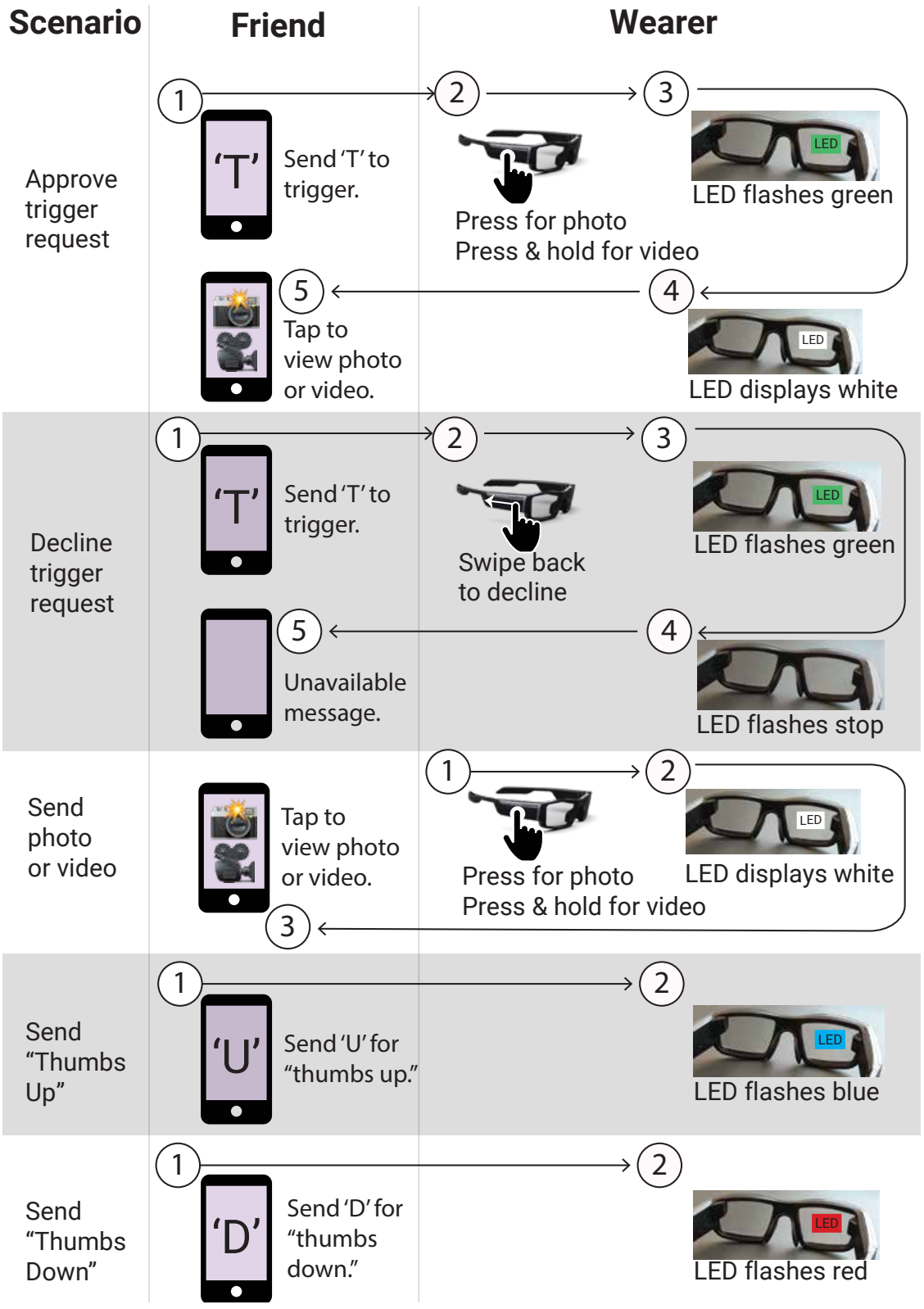

Fig. 4. Interaction flow between Wearer and Friend. This sequence diagram shows the input gestures, LED signals, and messaging codes that Friendscope uses. The Wearer can approve trigger requests and initiate photos/videos themselves using the same gesture. The Wearer's inward-facing LED flashes green, blue, or red to represent the Friend's trigger requests ('T'), "thumbs up" messages ('U'), and "thumbs down" messages ('D'), respectively.

- `MANUAL APPROVE MODE`: This mode declines all trigger requests that time out, requiring the Wearer to manually "approve" requests. This mode strikes a balance between giving Wearers control and giving Friends access to the Wearer's experience.

Auto Approve and Manual Approve modes behave identically while a trigger request is pending; the only difference between them is what happens when the request times out after 10 seconds. The Wearer can decline trigger requests in both of these modes. In that case, for the Friend it will appear as if the trigger requests timed out.

During a session, the Wearer can switch between modes at any time and initiate photos/videos themselves in any mode. Switching modes triggers a message to the Friend, informing them of the new mode, since this may influence the Friend's behavior and expectations.

## 3.2 Wearer Side: Camera Glasses App

The camera glasses app allows the Wearer to capture photos and videos. It also mediates trigger requests from the Friend. Captured videos (with audio) are ten seconds long to match the functionality of Spectacles [5].

Due to lack of a public API for current camera glasses [66, 78, 79], we prototyped Friendscope using a Unity [71] app that runs on Vuzix Blade [73] smartglasses. We chose Vuzix over other Unity-compatible products such as HoloLens [48] because the Vuzix are relatively lightweight and are the closest approximation of minimal camera glass interfaces.

Ideally, we would have tested the value of the shared camera by implementing Friendscope on current commercial camera glasses hardware such as Snapchat Spectacles or zShades rather than the Vuzix Blade. However, existing screenless commercial camera glasses lack public APIs. Because of that, we mimicked these camera glasses' functionality on the Vuzix Blade hardware to foster realistic scenarios and usage behaviors in our studies: our pilot studies, main study, and field exploration with 82 participants in total. Recall that camera glasses feature only a camera, a microphone, a button, and an inward-facing LED. Vuzix Blade smartglasses, by contrast, have many more hardware features. As a result, we explicitly disregarded much of the Vuzix Blade's hardware to simulate camera glasses. We repurposed its screen to represent a single mock LED (see third column of Figure 4), and we repurposed its touchpad to represent a simple button.

*3.2.1 Input actions.* The Wearer can perform three actions on their camera glasses: send a photo (on their own or while responding to a trigger request), send a ten-second video (on their own or while responding to a trigger request), and decline a trigger request. In general, the camera glasses need only a single button for the Wearer to perform these actions since they can be mapped to different gestures. Our prototype's Vuzix Blade camera glasses feature a touchpad on their right side, so we mapped the actions to the following gestures: a "press" to send photo, a "press and hold" to send video, and a "swipe back" to decline a trigger request (see Figure 4). Explicitly declining a trigger request sends an "unavailable" message to the Friend after 10 seconds as if the trigger request had timed out.

*3.2.2 Fast fulfillment of trigger requests.* Since fulfilling the Friend's trigger requests quickly can make Friends feel more tightly connected to the Wearer's experience, Friendscope actively creates opportunities for fulfilling trigger requests immediately. To do this, Friendscope holds all photos/videos initiated by the Wearer for 10 seconds before delivering them to the Friend. If the Friend sends a trigger request during this holding period, Friendscope sends the held photo/video immediately to create an instant response to that trigger request.

Under perfect network conditions, a naively-designed trigger request could take up to 25 seconds for a video to return (= 10 s "trigger request pending" state + 10 s to record + 5 s to transmit) and 12 seconds for a photo to return (= 10 s "trigger request pending" state + 1 s to capture + 1 s to transmit). Using the rapid fulfillment technique described above, Friendscope narrows this gap to only 5 seconds for a video and 1 second for a photo, keeping just the time needed to transmit the video or photo.

*3.2.3 Screenless LED interface.* To keep power costs low and make Friendscope applicable to as many types of camera glasses as possible, Friendscope's camera glasses app is screenless and uses a single LED as its display to the Wearer. As a result, we opted not to fully utilize the Vuzix Blade's built-in screen for our Friendscope implementation. Instead, as Figure 4 shows, we display just a colored area on the Vuzix Blade's screen to represent a "mock" inward-facing LED. This "LED" is usually off but turns on to communicate the following:

- It flashes green for ten seconds when a trigger request is received and is awaiting action.

- It displays white while a photo is taken or a ten-second video is recording.
- It flashes white when a photo or video is successfully sent.
- It flashes blue to show a Friend's "thumbs up" message.
- It flashes red to show a Friend's "thumbs down" message.

### 3.3 Friend Side: Messaging App

We created a basic text messaging app for our Friendscope implementation to demonstrate how Friendscope would work within a standard messaging app. The reason is that evidence suggests that Friendscope will work best for Friends if it is compatible with existing messaging apps and embedded within them rather than being a separate smartphone app. Most in-the-moment photo/video sharing today happens on instant messaging apps such as iMessage, WhatsApp [29], Facebook Messenger [14], and Snapchat [65], where people send and receive photos/videos seamlessly and spontaneously as part of their normal communication with each other [1]. These apps allow users to interact asynchronously and do not require them to coordinate with each other or be online at the same time. They also notify recipients whenever photos/videos are sent so that they can go online and join the conversation if they are not online already.

The Friend can send trigger requests by texting the Wearer a "T" in the messaging app, as shown in the right side of Figures 3 and 4. After sending the "T," they will get an automatic *"Trigger received!"* confirmation message. If the trigger request is approved, the Friend receives a series of messages to keep them updated about the status of their request, starting with a message saying, *"Trigger approved! Hold tight for a [photo/video]!"*, followed by *"Video/photo is being transmitted!"* then the countdown (*"5. . . "*, *"4. . . "*, etc.), and finally the photo or video.

If the trigger request is declined or times out after 10 seconds, they will get a message saying that the Wearer was unavailable. The "unavailable" message comes after 10 seconds in either situation, in order to afford plausible deniability [53]. Friends also get a notification message when the "sharing mode" is changed by the Wearer. For example, changing the mode to Auto Approve would display the following message to the Friend: *"Starting AUTO APPROVE mode now! Send 'T' to trigger a photo/video of what I'm seeing right now."* Likewise, changing the mode to Shared Camera Off would display: *"Starting SHARED CAMERA OFF mode now. Pausing trigger requests for a bit, but I'll keep sending you photos/videos whenever possible."*

During our pilot study with 22 participants, Friends requested additional ways to interact with Wearers. (Wearers can already communicate to their Friend by simply speaking while capturing a video). As a result, we made it possible for Friends to send "thumbs up" and "thumbs down" messages to Wearers by texting a "U" or a "D" (respectively) in the messaging app. These then flash the Wearer's LED blue or red, respectively. Liu et al. [42] and Kaye et al. [35] show that even extremely limited communication channels can be popular and can fulfill users' need for connection with friends.

## 4 EVALUATION

Friendscope's key design contribution is its shared camera, which enables in-the-moment experience sharing on camera glasses. With Friendscope, we wanted to understand how different shared camera configurations — Auto Approve, Manual Approve, and Shared Camera Off — lead to different types of experiences for users in terms of *fellowship* between users, *privacy* for the Wearer, and *sense of control* for both Wearer and Friend. We also studied how Friendscope's screenless design impacted users' experience. Shared Camera Off represents our baseline condition, with no shared camera at all (no trigger requests — only the Wearer can initiate photos/videos). This baseline allows us to understand the affordances and behaviors that the shared camera and trigger requests bring.

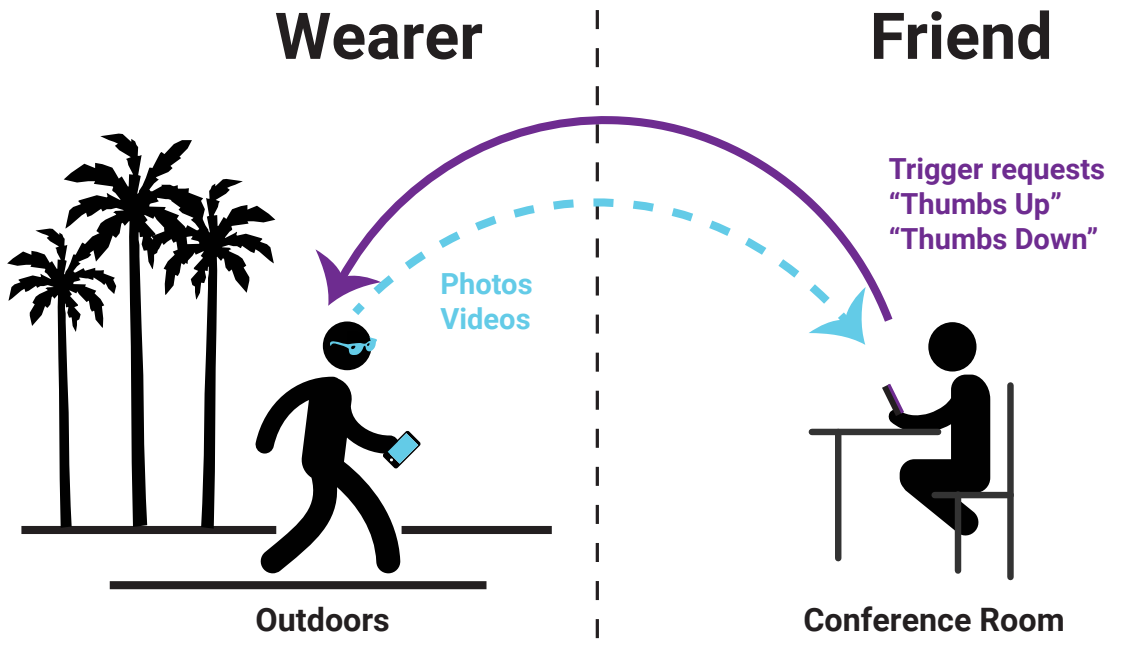

Fig. 5. User study scenario. The Wearer visited an outdoor area while the Friend stayed behind in a conference room. We ran the study in three tourist locations: Venice Beach in Los Angeles, Pike Place Market in Seattle, and Times Square in New York. Each Wearer/Friend pair went through Friendscope's three sharing modes (Auto Approve, Manual Approve, and Shared Camera Off) and answered surveys about their experiences with them.

## 4.1 Participants

Our main user study had 24 pairs of participants (48 participants total: 22 female and 26 male), but we ran pilot studies with 11 additional pairs of participants (22 participants total) first as part of our iterative design process. We recruited all 70 of our participants from a technology company using their employee mailing list. We recruited from ten diverse teams within the company, representing a wide variety of technical and non-technical roles including HR and sales/marketing. We recruited from three company offices: New York, Los Angeles, and Seattle. We asked people to participate in the study with their closest friend within the company.

In our main study, eight participants self-reported as having extensive experience with wearable camera interfaces such as GoPro [19], Google Glass [18], or Spectacles [66]. Participants' ages ranged from 18 to 44: 13 aged 18–24, 25 aged 25–34, and 10 aged 35–44.

## 4.2 Study Procedure

Our study used a within-subjects design in which each pair of participants experienced all three modes of Friendscope together. Within each pair, we randomly assigned a Wearer and a Friend unless one of the participants required glasses. Since our Vuzix Blade glasses did not have prescription lenses, we assigned participants who needed prescription lenses to be the Friend.

We were careful to choose a realistic scenario during our main study. We took the Wearer to a popular tourist destination nearby to explore it freely as a tourist would and had the Friend stay behind in a conference room to interact with the Wearer through trigger requests and thumbs up/down messages. The popular destinations were Times Square, Venice Beach, and Pike Place Market for New York, Los Angeles, and Seattle participants, respectively. Participants were given the chance to practice using Friendscope before splitting up. Figure 5 illustrates our study design.

Each participant pair tried Friendscope's three shared camera configurations (*Shared Camera Off*, *Auto Approve*, and *Manual Approve*) in a counterbalanced order, where Shared Camera Off represents our baseline condition. They tried each sharing mode for 10 to 15 minutes. After trying each mode, we asked each pair to complete a questionnaire about their experience with that mode (see below for questionnaire details). Once they had tried all sharing modes, we reunited the participants and asked them to complete a final survey comparing the three modes. The sharing mode was our independent variable, and the user responses, including ratings of fellowship, control, and privacy are our dependent variables. We concluded by performing a semi-structured interview with each pair to gather their final thoughts and takeaways. As we describe in Section 4.4, questions in the interview probed their prior experience with live-streaming, asked them to compare and contrast the modes, and unearthed details about the screenless experience.

In summary, each main study participant completed five questionnaires during the main study in addition to the semi-structured interview at the very end. The five questionnaires were the pre-study questionnaire, one questionnaire for each of Friendscope's three sharing modes (right after they tried out that mode), and one final questionnaire. Each main study session took 2 hours and 15 minutes in total.

### 4.3 Survey Design

The pre-study questionnaire collected simple demographic information, including age and gender as well as prior experience with camera glasses and smartglasses (including Google Glass [18], Spectacles [12], etc.). We also collected self-reports of levels of privacy concern, using relevant questions from the "Global Information Privacy Concern" portion of the IUIPC privacy questionnaire [44]. We used the single-item, pictorial Inclusion of the Other in the Self scale to assess interpersonal closeness in both the pre- and post-study questionnaires [3].

We constructed the three sharing mode questionnaires in three primary steps. First, we held a team-wide brainstorming session with all authors of this paper to discuss the potential costs, benefits, and societal ramifications of shared cameras. This discussion brought forward several topics such as closeness between friends, togetherness, enjoyment, privacy, sense of control, control over access, social concerns, and others.

Second, we drafted a survey and conducted a set of pilot studies, in which we had participants use Friendscope outdoors for a short period of time and then complete a questionnaire asking about their impressions with respect to the positive and negative impacts that we identified. Participants' responses to many of these questions were similar, indicating that our initial set of possible impacts of shared cameras could be merged into a smaller set of broader themes. Based on the findings from our pilot studies, we arrived at three broad themes that form the structure of our surveys — fellowship, sense of control, and privacy.

Third, for each theme, we crafted parallel questions for Wearers and Friends addressing each aspect of using Friendscope. For example, to understand how the camera modes affected fellowship: *"I felt closer to my friend by giving them the ability to trigger my camera"* (for the Wearer) and *"I felt closer to my friend because I had the ability to trigger their camera"* (for the Friend). To probe our system's effect on participants' sense of control, we asked these two questions in addition to others: *"I felt that I always knew when photos/videos were being taken"* (Wearer) and *"I felt as if I took the photos/videos that I received"* (Friend). To address the theme of privacy, a subset of our questions includes: *"I was never worried about sharing something I didn't want to"* (Wearer) and *"I never felt that I was invading the privacy of my friend"* (Friend).

We asked questions like the above for each sharing mode (Manual, Auto, Off). The surveys featured both open-ended responses and 7-point Likert scale ratings, where '1' indicated Strongly Disagree and '7' indicated Strongly Agree. This enabled us to understand the experience for both sets of users (Wearer and Friend) across all sharing modes in detail. The final questionnaire asked participants to rank each mode in terms of overall preference (see Figure 6), fellowship (see Section 5.2), privacy (see Section 5.4), and feelings of control (see Section 5.3). We additionally used the NASA TLX [24, 25] to ask about cognitive load and mental effort while using the system.

### 4.4 Semi-Structured Interview

At the end of every study session, after each participant had completed their final questionnaire, we engaged each pair of participants in a semi-structured interview to further elicit details about the experience. The topics we discussed in the interviews included concepts related to Fellowship (including social presence, connectedness, interaction, immersion, and closeness), Privacy (intrusiveness, vulnerability), and Control (access, risk). Here we share a subset of our interview guide which included the following questions, following best practices for conducting semi-structured interviews [8, 69, 77]: *"Can you describe the experience of letting your camera be shared?"*, *"What were your main concerns when using this system?"*, *"Can you go into detail about how you felt when receiving / sending a trigger?"*, *"Compare and contrast the different modes (why did you choose this ranking?)"*, *"What did receiving/sending a trigger mean to you — how did you interpret that?"*, *"Can you talk about any concerns you had about seeing or sharing something you didn't want to?"*, *"How did the lack of ability to preview affect your experience?"*. We also asked participants to directly compare 1:1 messaging, group sharing, livestreaming, and Friendscope.

Following best practices for Charmaz's grounded theory [8], we simultaneously engaged in analysis and data collection, iteratively constructing our analytic frame and updating our question prompts for future interviews as we synthesized and identified emerging themes. We read and analyzed all interview data and discussed all emerging themes [47], which are presented below.

## 5 STUDY RESULTS

Our study data includes survey responses, usage logs, and qualitative feedback through our open-ended survey questions and semi-structured interviews. We analyzed quantitative survey responses using non-parametric Friedman tests[1] with an alpha level of 0.05. Additionally, our post-hoc tests used Wilcox tests with Holm-Bonferroni method for correcting multiple comparisons (all reported $p$-values are adjusted).

For our thematic analysis process, we first transcribed each semi-structured interview, then performed open-coding [68] on the transcripts. Following best practices for grounded theory [8], we simultaneously engaged in analysis and data collection, iteratively constructing our analytic frame and updating our question prompts for future interviews as we synthesized and identified emerging themes. We iteratively reviewed and refined these into a closed set of codes, which we then re-applied to the transcripts as we performed additional interviews. Since the inception of grounded theory, it has split into three main branches: Strauss and Corbin, Glaser, and Charmaz [61]. We embrace Charmaz's constructionist research style that understands knowledge as co-constructed between interviewee and researcher [8, 9]. Our analysis is interpretivist, seeking to understand how our informants create meaning in their experiences [17], and is rooted in the social construction of knowledge and polysemic understandings of truth [40]. Since we followed a grounded theory approach, it was not appropriate or necessary to compute inter-rater reliability [47].

Here we report our findings about how different *shared camera* configurations — Auto Approve, Manual Approve, and Shared Camera Off — lead to different types of experiences for users in terms of *fellowship* between users, *privacy* for the Wearer, and *sense of control* for both Wearer and Friend. We split our discussion into each of these topics, and further split it between Wearers' and Friends' perspectives.

### 5.1 General Findings

Wearers sent a total of 864 messages during the study: 389 photos and 414 videos. Remote Friends sent 631 messages: 242 trigger requests, 263 thumbs up messages, and 126 thumbs down messages.

Towards the end of the study, we asked participants to share their overall feedback. Both Wearers (*Md=6.5*) and Friends (*Md=6*) rated Friendscope as being fun to use on a 7-point Likert scale. Wearers rated it as requiring low mental effort (*Md=2*). When asked whether they felt Friendscope would add value to existing camera glasses products, both Wearers (*Md=7*) and Friends (*Md=6*) rated the system very highly.

[1]While opinions on the importance of normality vary dramatically, and statisticians continue to debate appropriate techniques [37], in this work we follow standard statistical analysis best practices. Specifically, we employ non-parametric statistical tests since the Shapiro-Wilk test showed that our data violated the Assumption of Normality.

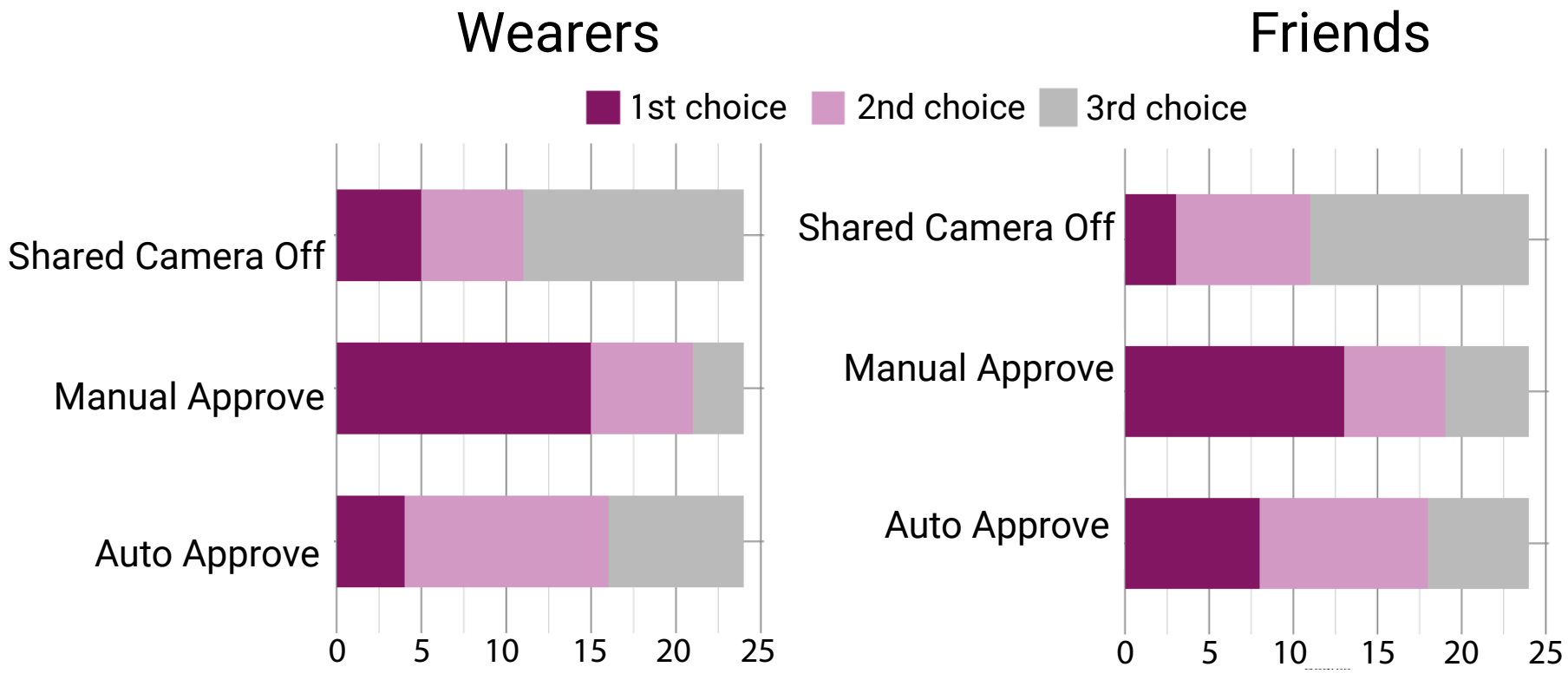

Fig. 6. Sharing mode preference rankings. Overall, Wearers (Left) and Friends (Right) both preferred Manual Approve mode. This preference was statistically significant for Wearers ($p < 0.01$).

We also asked participants to rank Friendscope's three shared camera configurations by order of preference. As shown in Figure 6, Wearers and Friends both preferred Manual Approve mode the most, possibly because this mode strikes a balance between giving Wearers control and giving Friends a feeling of liveness. This preference was statistically significant for Wearers ($\chi^2(2, 24) = 10.75, p < 0.01$).

## 5.2 Findings: Fellowship

Here we report our findings related to fellowship: how different shared camera configurations give users a sense of being together with each other.

*5.2.1 Fellowship for Wearers.* As shown in Figure 7a, we found a significant main effect of sharing mode on Wearers' feeling of togetherness: $\chi^2(2, 24) = 11.39, p < 0.001$. Post-hoc analyses revealed that Wearers ranked both Auto Approve (*Md=5*) and Manual Approve (*Md=5*) sharing modes as fostering a significantly higher ($p < 0.01$) feeling of togetherness than Shared Camera Off (*Md=3.5*). Both Manual Approve and Auto Approve allowed Friends to send trigger requests, while Shared Camera Off did not. Trigger requests acted as signals to the Wearer that their Friend was watching live, making them feel more together with each other.

In addition, we found a significant main effect on how close the Wearer felt with their Friend after sending photos/videos in response to trigger requests compared to sending photos/videos on their own: $\chi^2(2, 24) = 13.13, p < 0.01$. Post-hoc analyses revealed that Wearers felt closer to their Friends after sending photos/videos in both Auto Approve (*Md=5*) and Manual Approve (*Md=5*) ($p < 0.01$) modes compared to Shared Camera Off (*Md=4*), with $p < 0.01$. Closeness ratings using the IOS scale [3] stayed consistently high for both Wearers and Friends after using our system.

During the semi-structured interview, one third of Wearers confirmed that Friendscope felt "more personal" than other forms of image-based communication such as sending photos over an instant messaging platform. In particular, Wearers emphasized the sense of being directly connected, in the moment, to their friend:

> *W19: It was nice to get that immediate feedback that someone is paying attention to what you're doing. The trigger also let you know when someone was engaged. It was cool to have someone watching what you were creating.*

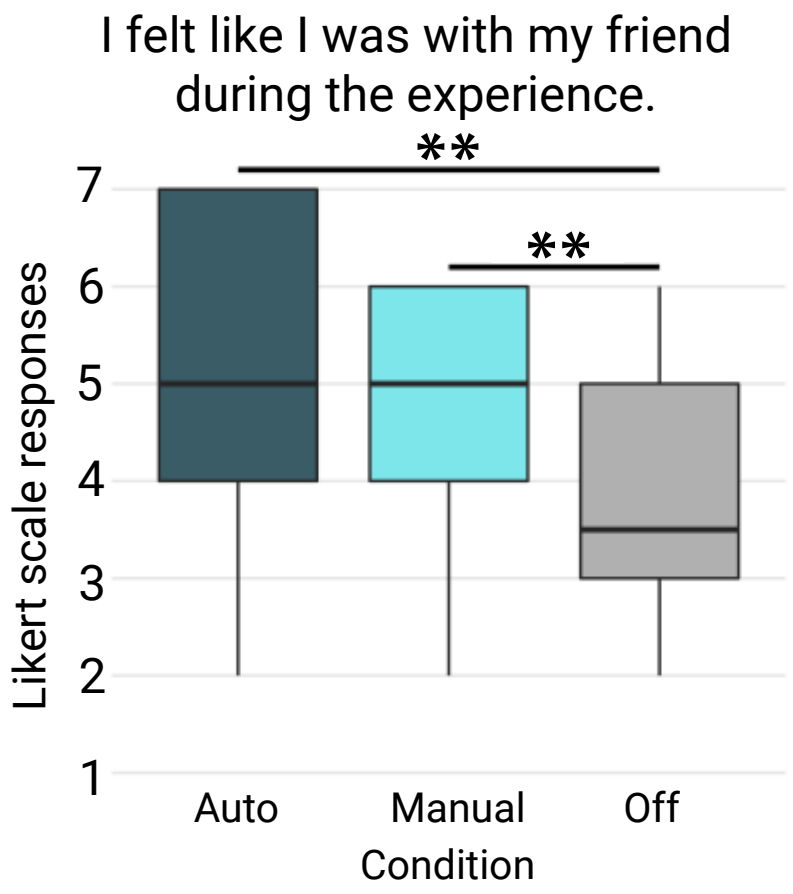

(a) Wearers

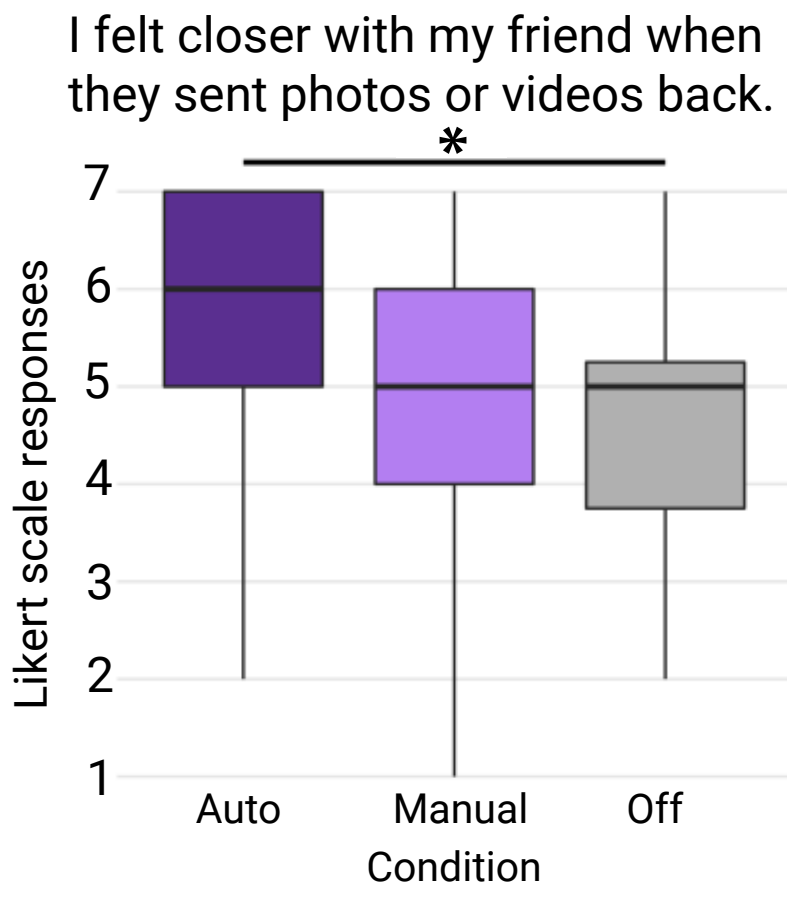

(b) Friends

Fig. 7. Fellowship ratings. (a) Wearers rated both Auto Approve and Manual Approve sharing modes as fostering a significantly higher ($p < 0.001$) feeling of togetherness than Shared Camera Off. (b) Friends rated Auto Approve significantly better than Shared Camera Off at making them feel closer to the Wearer.

Wearers favorably compared Friendscope to a live video call:

*W22: I felt like I was streaming something to him even though that's not actually what's happening.*

Throughout the interviews, Wearers were enthusiastic about the idea that their Friend was "jumping into [their] head" (W11) and actively watching their shared images right at that moment. The idea that someone was paying attention seemed to increase the sense of togetherness and closeness for Wearers. In other words, the "liveness" experienced by both Wearers and Friends was associated with the "sharing event" of the photos or videos (as opposed to the original event). Shared attention has been shown to increase "mood infusion" and emotional experiences [62–64]. Our findings resonate with Haimson and Tang, showing that the immediacy of using Friendscope made for an engaging experience [22].

*5.2.2 Fellowship for Friends.* As shown in Figure 7b, we found a significant main effect of sharing mode on how close Friends felt with Wearers when receiving photos or videos ($\chi^2(2, 24) = 11.39, p < 0.001$). Further analysis revealed that Friends rated Auto Approve (*Md=6*) significantly better than Shared Camera Off (*Md=5*) at making them feel closer to Wearers. Note that Shared Camera Off represents our baseline condition where there is no shared camera at all and the Wearer must initiate all photos/videos. Being able to trigger the Wearer's camera seems to have made Friends feel closer, with the system described as "a step above existing communication" and "way more personal."

In addition, in our interviews, Friends reported feeling a live connection with the Wearer even though Friendscope itself is asynchronous:

*F16: And since it was live – it wasn't taken a while ago – it was happening right there, it felt connected.*

This feeling of liveness made the shared experience more intimate, interactive, and personal for Friends:

*F4: I did like the fact that [my friend] would talk to the camera so it felt very close. I was experiencing it much more than just viewing [it].*

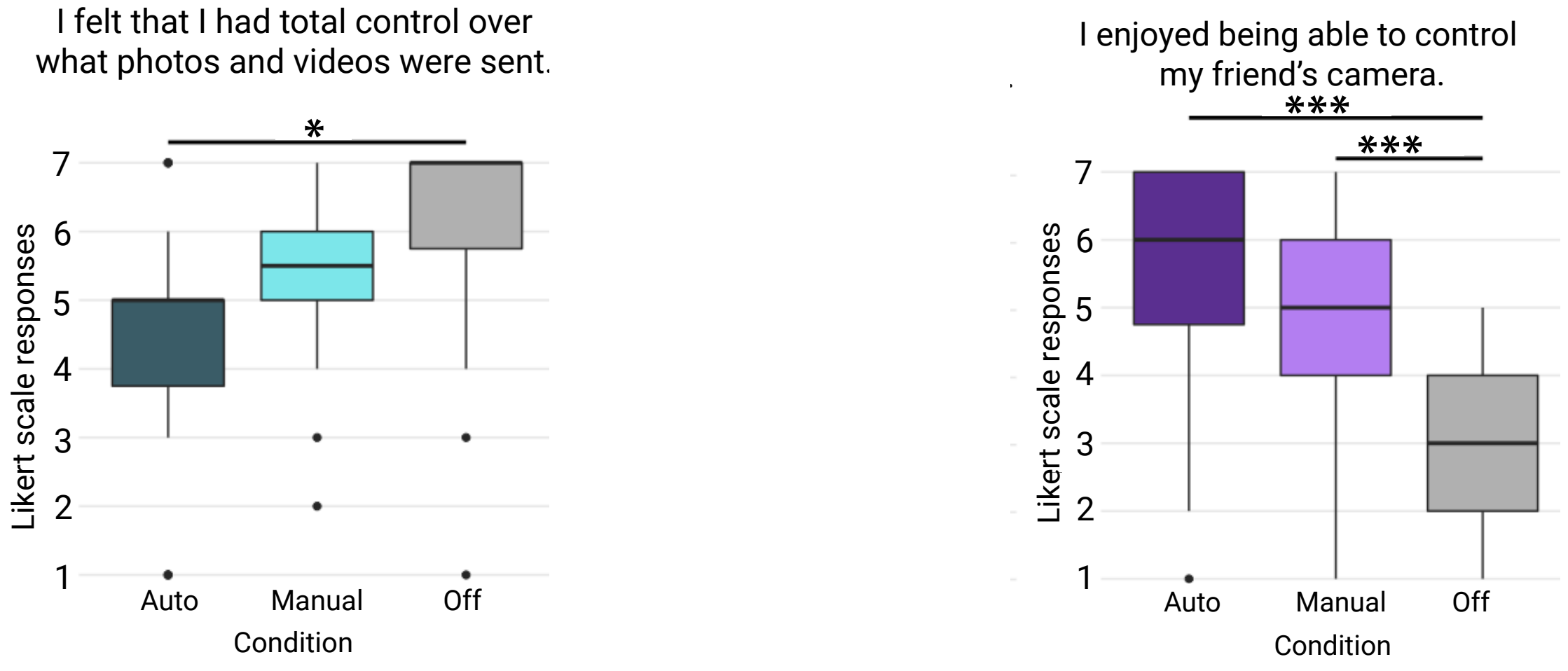

(a) Wearers (b) Friends

Fig. 8. Control ratings. (a) For Wearers, post-hoc analysis found a higher feeling of control ($p < 0.05$) in Shared Camera Off mode than in Auto Approve mode. There was no statistically significant difference between Shared Camera Off and Manual Approve modes. (b) Friends felt more able to control the Wearer's camera ($p < 0.01$) in Auto Approve and Manual Approve modes than they did in Shared Camera Off mode.

Many participants directly compared the experience to live-streaming or watching a video, which they had all experienced:

*F3: [I was] more participating than watching a video.*

Several Friends described the shared camera as an embodied experience and felt that they were "sharing [the Wearer's] eyes." Friends also described the experience as a more intimate and personal form of livestreaming. Even those who do not normally watch livestreams found the shared camera appealing:

*F3: Livestreaming is more interesting when it's just for you.*

This resonates with earlier work on videos filmed from a first-person view (FPV). Footage from this perspective has been shown to increase the sense of 'seeing with the eyes of another' [45, 56]. Our work builds on these findings, emphasizing the unique potential for camera glasses to enable closeness, making them uniquely suited to sharing intimate moments with close friends and family. Friendscope's system design also takes advantage of known benefits of photo- and video-based sharing, which afford greater opportunities for self-expression than text-based systems [75] and enable more frequent communications with close friends and family [70] (see [1] for a thorough overview).

A lack of immediacy and intimacy can reduce social presence when people communicate through mediated channels [20, 21]. Promoting social presence should promote feelings of connectedness [28]. The shared camera's focus on immediacy and intimacy enhanced feelings of social presence and connectedness.

### 5.3 Findings: Sense of Control

Here we report our findings on how the shared camera affects Wearers' and Friends' sense of control over what is captured. Figure 8 shows our survey results.

*5.3.1 Sense of Control for Wearers.* As Figure 8a shows, we found a significant main effect of sharing mode on Wearers' feeling of control over what was captured: $\chi^2(2, 24) = 31.75, p < 0.0001$.

Post-hoc analysis revealed a higher feeling of control ($p < 0.05$) in Shared Camera Off mode (*Md=7*) compared with Auto Approve mode (*Md=5*). Wearers preferred Manual Approve mode the most:

> *W28: I liked Manual Approve mode the most. That was the perfect blend of spontaneity and having me in control.*

*5.3.2 Sense of Control for Friends.* As we expected, Friends ranked both Manual Approve and Auto Approve modes as providing them greater control compared with Shared Camera Off, the baseline condition where there is no shared camera at all. Friends "felt free to control [the Wearer's] camera" ($p < 0.01$) to a greater extent in Manual Approve (*Md=5*) and Auto Approve (*Md=5*) modes compared with Shared Camera Off mode (*Md=2*). As Figure 8b shows, Friends enjoyed being able to control the shared camera significantly more ($p < 0.001$) in Manual Approve (*Md=5*) and Auto Approve (*Md=6*) modes as compared with Shared Camera Off mode (*Md=3*). We did not find a significant difference between the extent that friends enjoyed Auto Approve mode and Manual Approve mode.

In addition, the need for a gesture in Manual Approve mode seemed to increase feelings of interactivity, consent, and connection:

> *F27: Manual mode felt like it was more interactive. With Shared Camera Off mode she's just shooting things to me. With Manual I felt like it's an actual interaction. The fact that there was a request and an approval element made me feel "Oh! I'm interacting with [my friend]." I'm sending you this, this is what you asked for and I approved: it really creates that connection and interaction.*

Friends were sensitive about how frequently they sent trigger requests in Auto Approve mode (*Md=5*), which they described as "spamming" during the interview. They did not want to cause the Wearer to share something they did not want to share. By contrast, Friends felt comfortable triggering the camera as frequently as they wanted to in Manual Approve mode (*Md=6*).

The practice of lifelogging represents an early exploration around the perception of a lack of control and privacy with body-worn cameras. Lifelogging involves a small camera clipped onto clothing or otherwise attached to the body that automatically captures first-person perspective photographs at a predetermined rate throughout the day. Lifelogging is typically automatically triggered, meaning the wearer has even less control than when using Friendscope's shared camera. Researchers found that users prefer to maintain control over what pictures are captured and shared. Kärkkäinen et al. [32] found that limiting who can see the images and automatically deleting older images both provide this sense of control. The shared camera design that we explore with Friendscope incorporates both of these techniques. Incorporating these has the additional benefits of avoiding known [39] issues with wearable cameras, including self-censoring behaviour by the user.

### 5.4 Findings: Privacy

For the concept of a shared camera to be embraced by users, it must accommodate their privacy preferences. Here we report our findings on what users thought of the different shared camera configurations from a privacy perspective. We report findings from users in general and from those who self-describe as being sensitive to privacy specifically.

As a means of gauging our participants' sensitivity to privacy concerns, we averaged their responses to relevant questions from the "Global Information Privacy Concern" portion of the IUIPC privacy questionnaire [44] to form a new "privacy awareness metric." We included questions regarding personal beliefs and behaviours around technology, and we omitted questions about unauthorized data usage and other topics not related to shared cameras. By this metric, our participants rated themselves as being moderately concerned with personal privacy on average: 4.7 out of 7.

*5.4.1 Privacy for Wearers.* We found a significant difference in how Wearers ranked the three sharing modes in terms of respecting privacy: $\chi^2(2, 24) = 24.33, p < 0.00001$. Wearers ranked the modes in the following order from most to least privacy-preserving: Shared Camera Off, Manual Approve, and Auto Approve. We also found that Wearers were significantly less "worried about [accidentally] sharing something they did not want to" ($p < 0.01$) in Shared Camera Off mode (*Md=6*) and Manual Approve mode (*Md=6*) compared to Auto Approve mode (*Md=4.5*). Additionally, Wearers felt equally "in control of [their] privacy" and privacy settings in each mode (Same scores for both questions: Shared Camera Off mode *Md=7*, Manual Approve mode *Md=6*, Auto Approve mode, *Md=5*).

Next, we examine a subset of our users who self-described as being particularly sensitive to personal privacy concerns. We define this subset to be users whose personal privacy awareness metric value (described above) was at least 6 out of 7. Five Wearers and five Friends matched this criteria.

Six of these ten participants ranked Manual Approve mode as their most-preferred mode, which means that even these privacy-sensitive participants preferred to use Friendscope's shared camera instead of turning it off (Shared Camera Off). Only two of these ten participants (one Wearer, and one Friend) preferred Shared Camera Off, and two preferred Auto Approve. Participants' comfort with the shared camera stemmed from them getting to choose who to include in their Friendscope session:

> *W8: In general I don't like to share much so all my [social media] is private. But with this I felt more comfortable because it was just an experience I was having with my friend.*

*5.4.2 Privacy for Friends.* By and large, Friends agreed that they "never felt that [they were] invading the [Wearers'] privacy," ranking all three sharing modes relatively highly for protecting Wearers' privacy. The ratings were *Md=6* for Shared Camera Off, *Md=5.5* for Manual Approve, and *Md=5* for Auto Approve modes. These differences were not statistically significant.

### 5.5 Findings: Screenless Experience

Friendscope's screenless design makes it compatible with consumer camera glasses which generally lack screens and speakers. This limits the types of interactions that are possible, including the ability to preview photos/videos before sending, and the ability to see Friends' messages. During our interviews, participants consistently reported their experience with Friendscope's screenless design. We summarize their remarks here.

*5.5.1 Screenless Experience for Wearers.* The lack of a screen meant that Wearers were not able to see text message responses from their Friends. While Wearers could take the phone out of their pocket to see messages from their Friends, Wearers specifically shared that they preferred to not be "connected to [their] phone the entire time". The lack of a screen also meant that Wearers were not able to preview their photos or videos before sharing them. 18 out of 24 Wearers were comfortable with not being able to preview photos/videos before sending, with six actually *preferring* not being able to preview. For the 18 Wearers, the lack of a preview helped them stay immersed in their activity and enhanced their feelings of sharing it live:

> *W16: No I don't want preview. That spoils my experience of the ongoing activity.*

> *W4: I really liked not being able to curate [the images]. I really felt like I was sharing them in the moment.*

The remaining six Wearers would have preferred having a way to preview and/or edit their photos/videos before sending, the main reason being to annotate them with personal context.

*5.5.2 Screenless Experience for Friends.* Since Friendscope is screenless, the only way for Wearers to converse with their Friend was to speak while recording videos. Wearers quickly figured this out, and Friends valued these types of videos very highly. During the semi-structured interview, one Friend described watching these videos as "experiencing more than just viewing." The narrations seemed to increase Friends' sense of being with the Wearer:

*F4: The way [the Wearer] was using it felt very intimate. I felt like I got to know [her] better.*

When Wearers did not narrate their videos, Friends described the videos as "generic," "random," or "impersonal." Friends also found photos less appealing than videos for this reason.

In general, throughout the interviews Friends expressed a strong desire to communicate with Wearers in more ways than what Friendscope provided. Even though Friendscope supported limited feedback via "thumbs up/down" messages, the majority of Friends (18/24) felt that the communication was one-sided. Since Friendscope was designed around a small number of hardware components, Friends could not ask Wearers questions or respond to Wearers as they would be able to if they used a smartphone:

*F25: It felt like she was trying to have a conversation and I was ignoring her.*

14 Friends expressed a desire to send emojis or other simple yet expressive messages beyond "thumbs up/down" messages. Several Wearer/Friend pairs reported that they had repurposed the "thumbs up/down" messages to mean other agreed-upon things. For example, one pair reported that the "thumbs up/down" messages were "practically an inside joke already" (F10). This resonates with earlier work by Liu et al. [42] and Kaye et al. [35], who show that even extremely limited communication channels can be popular and can fulfill users' need for connection with friends.

## 6 FIELD EXPLORATION

We conducted a small field exploration in addition to our main study to get a short glimpse of how people would use shared cameras "in the wild." We report our findings here as basic observations only and not as a complete field study. We loaned camera glasses to six interested users for longer periods of time (from one day up to one week) and allowed them to use Friendscope as they saw fit. These users were from the same technology company as our previous study. They acted as Wearers and chose a Friend, resulting in six pairs of users (12 field exploration participants total).

Five of these users were female, 3 were male, and 2 preferred not to respond. Seven were aged 25–34, one was aged 35–44, and four preferred not to respond. As with the main study, these users represented a wide variety of roles within the company, including many non-technical roles. We gave these participants the same onboarding procedure as in the main study, but we encouraged them to use Friendscope whenever and however they wanted, including changing the sharing mode (shared camera configuration) as they saw fit. We performed a semi-structured interview with each pair collectively after they finished trying out Friendscope.

Wearers used Friendscope in many contexts: both indoors and outdoors, for both daily life and special events, and both in and out of coordination with their respective Friends. We describe some of the use cases and users' resulting experiences here.

One pair of participants borrowed the glasses for a day. The Wearer wore the glasses during a children's baseball game while his Friend stayed at his office. The Friend enjoyed the immediate sharing, saying that it greatly increased his sense of being with the Wearer:

*F1: I loved using the trigger function to feel like I was live participating...It made me feel like I was living the experience with him.*

This pair had such a positive experience that they asked to borrow the glasses every week afterward. Their experiences echo that of our main study participants, who also found that Friendscope makes them feel together with each other in the moment.

Another Wearer used the glasses to share a personal celebration at home with her Friend. She emphasized that the immediacy of the connection increased her sense of being with her Friend:

*W3: It was really cool because when I was sharing that with her, she immediately responded...I really felt like she was there even though she wasn't.*

The Friend emphasized the special experience of having a trigger request approved:

*F3: It was like I was being allowed into her world. Her saying yes to [my trigger] request is like opening a treasure box.*

Another pair typically works together in the same office, but the Friend was working remotely when the Wearer borrowed the glasses. Both Wearer and Friend spent several hours going about their normal activities while using Friendscope. They always used Auto Approve mode. The Wearer interpreted trigger requests as confirmations that her Friend saw her message at that moment, finding that feedback helpful and not requiring "thumbs up/down" messages from the Friend.

## 7 DISCUSSION AND LIMITATIONS

Our results show how different shared camera configurations — Auto Approve, Manual Approve, and Shared Camera Off — lead to different experiences and affordances for users in terms of fellowship, privacy, and sense of control. We summarize our results and present our design recommendations for future experience sharing systems in Table 1. In this section, we discuss these recommendations, Friendscope's privacy implications, and our study's limitations.

### 7.1 Continuous access supports in-the-moment experience sharing

Experience sharing systems have traditionally relied on continuous streaming to create a feeling of liveness for remote viewers, but with some major trade-offs. Continuous streaming is too hardware-intensive to last long on consumer camera glasses, and it requires consistently high network bandwidth for acceptable video quality.

Our findings suggest that the concept of a shared camera makes it possible to avoid these limitations. By providing continuous *access* rather than continuous streaming, the concept of a shared camera enables users to share experiences in the moment using camera glasses. Users emphasized the sense of being directly connected — live — to each other when using Friendscope and described it as a more intimate form of livestreaming. Trigger requests signaled to both Wearers and Friends that their partners were connected "live." Even users who do not normally watch livestreams found Friendscope appealing.

In traditional livestreaming, the streamer initiates the session and streams their experiences continuously, involving them and their viewers throughout the session. Friendscope's shared camera design, by comparison, gives both parties more agency and flexibility. It allows the Wearer to share their highlights whenever they want rather than continuously, and it allows the Friend to learn about the Wearers' state and send trigger requests whenever they want to see what the Wearer is up to.

Building on this, future livestreaming systems can also incorporate the concept of continuous access in addition to (or instead of) continuous viewing. They might, for example, allow the streamer to "mark" or "highlight" interesting moments as they are happening, then notify friends who are not yet watching about those opportunities to tune in. Those friends, in turn, could send trigger requests to catch up on those interesting moments — in the moment — without having to follow the entire livestreaming session. Future continuous access-based systems can also allow the Wearer to add their status with the session invite to indicate what they are up to, i.e., their state, or prompt the Wearer to capture photos/videos at regular intervals to keep the Friend abreast of what they are doing throughout the session.

Table 1. Summary of our study findings and design recommendations for experience sharing systems.

| Theme | Study Findings | Recommendations |
|---|---|---|
| Fellowship | Users emphasized the sense of being directly connected — live — to each other when using Friendscope and described it as a more intimate form of livestreaming. | While *continuous streaming* is a popular approach for sharing live experiences, consider supporting *continuous access* instead or in addition. The mere feeling of having access can promote the feeling of closeness. |
| | Lightweight signals such as "trigger requests" signaled to both Wearers and Friends that their partners were connected "live." | Even very minimal interaction techniques can provide feelings of immediacy and intimacy in experience sharing systems. Consider incorporating "trigger requests" and other forms of lightweight signals in addition to standard message types such as text and voice notes. |
| Privacy | 1:1 sharing enhanced participants' comfort with the shared camera. | To mitigate privacy concerns, consider allowing users to choose small, private groups to share content with. |
| | Manually approving trigger requests assuaged Wearers' concerns about sending content accidentally. | Designing interactivity into a system can help communicate users' intent. Consider enabling explicit actions to signal explicit consent. |
| Control | Both Friends and Wearers preferred Wearers to have complete control over what was shared. | Consider using multiple, complementary techniques to give Wearers control –– from allowing them to invite selected friends only, to switch sharing modes anytime, and to reject or approve each trigger request. |
| Screenlessness | Many participants preferred the spontaneity of having no screen. | Consider embracing screenlessness. Not having a preview can support spontaneous, low-stakes communication by preventing users from polishing their output. |
| | Friends who wished the system had a screen wanted to use it for mostly low-fidelity responses. | Bidirectional interaction between friends is important, but consider keeping things simple even if the form factor allows higher fidelity. For instance, emojis or icons may be sufficient or preferred over text or photo responses for many. |

These findings reinforce Neustaedter et al.'s recent finding [54] that it is "the *ability* to see [a friend's video] rather than the *act* of seeing [the video constantly] that ma[kes] video a powerful connector." The concept of continuous access is generalizable and can be applied to existing or new systems regardless of hardware or software limitations. We hope that Table 1 acts as a catalyst for researchers and designers to explore the design space enabled by the concept of shared cameras further.

### 7.2 Shared cameras do not feel privacy-invasive for users when the owner has control

The idea of sharing control of one's personal camera may seem invasive to privacy. Friendscope helps Wearers maintain their privacy in multiple ways: by allowing them to choose who to include in their session, by supporting ephemeral messaging, and by giving them control to approve or decline each trigger request regardless of sharing mode. Our findings suggest that these strategies work.

Of Friendscope's three shared camera configurations, including Shared Camera Off (our baseline condition), both Wearers and Friends preferred Manual Approve mode the most in our forced ranking (Figure 6). To them, Manual Approve struck the right balance between providing access to the shared camera and maintaining the Wearer's control over what is shared. When we designed Friendscope, we hypothesized that Friends would have a clear preference for Auto Approve mode since that gives them the most access to the Wearer's camera, but that was not the case. Friends enjoyed the interactivity of the Wearer manually approving their trigger requests, and they also enjoyed knowing that with Manual Approve mode they would not cause the Wearer to inadvertently send something they do not want to.

Even privacy-sensitive users preferred to use the shared camera in Manual Approve mode vs. not having a shared camera at all (Shared Camera Off), primarily because they got to choose who to include in their Friendscope session. Hence, we believe that it is crucial for any future shared camera system to include a Manual Approve mode.

Manual Approve mode was not only the most preferred mode in our studies but it is also the mode that can prevent misuses of Friendscope-like systems such as remote "spying" and invading bystanders' privacy. Manual Approve mode's design grants full control to the Wearer and maintains existing norms around camera glasses, specifically: (1) an externally visible LED to inform bystanders when recording is occurring, and (2) a visible hand gesture when capturing or approving a photo/video. The ephemeral nature of all photos and videos captured by Friendscope goes beyond these existing privacy-protecting expectations to help additionally protect the privacy of people around the Wearer. Note that social norms around technology do evolve, and researchers continue to track expectations around data glasses in particular [38]. In the future, features like auto-expiration of sessions can make such systems even more privacy-protecting.

### 7.3 Camera glasses do not need screens to enable interaction between friends

Screens might seem necessary to allow the Wearer and Friend to interact with each other, but our findings show that even simple indicators such as LEDs can serve that purpose. Additionally, screenless designs make experience sharing systems such as Friendscope more glasses-friendly. Many camera glasses do not have screens, and screens are power-intensive. During our study, many participants preferred the spontaneity of having no screen, and those who wanted a screen only wanted to show simple messages such as text or emojis.

We believe that future experience sharing systems can leverage more sophisticated LEDs, dot matrix displays, or even audio to allow the Wearer and Friend to interact with each other and still be glasses-friendly. Future systems can even incorporate physiological data such as biosignals since they have been shown to be powerful for building social connection [42, 76].

### 7.4 The concept of a shared camera is designed for camera glasses, but can be generalized to a wide variety of smart glasses and wearable camera interfaces

We designed Friendscope to make in-the-moment, interactive experience sharing possible on camera glasses, which have a very limited set of hardware (not even a screen!) and support only a delayed form of sharing. We believe, however, that Friendscope's design concepts can benefit larger devices such as smartglasses and other wearable camera interfaces as well, even if it is implemented exactly as we have done in this paper.

For example, if the Wearer and Friend want to participate in an experience together but do not want to be online the entire time, they could use Friendscope to have a "lighter" form of interactive connection compared to a continuous livestream or video call. Neustaedter et al. found that users prefer *not* to have to attend a friend's or family member's live video call during their entire experience [54]. In addition, a user could use a Friendscope feature instead of a livestream or video call to save on battery life or counter poor network connectivity. Note that the current implementation of Friendscope does not rely on hardware such as speakers or screens, but using these on smart glasses could enable a richer experience. A similar "shared camera" experience is supported by the Clos application [72], which allows photographers to take pictures remotely [16].

## 8 CONCLUSION

In this paper we presented Friendscope, an experience sharing system for camera glasses that enables in-the-moment, interactive experience sharing on lightweight consumer camera glasses. Through our pilot study, main user study, and field exploration with 82 total participants, we investigated how different shared camera configurations — Auto Approve, Manual Approve, and Shared Camera Off — lead to different types of experiences for users in terms of *fellowship* between users, *privacy* for the Wearer, and *sense of control* for both Wearer and Friend. We found that both Wearers and Friends emphasized the sense of being directly connected to each other — in the moment — when using the shared camera, and described it as a more intimate form of livestreaming. We also found that shared camera systems should employ a Manual Approve mode as Friendscope does; even privacy-sensitive users preferred Manual Approve over not having a shared camera at all and were comfortable using Friendscope in its Manual Approve mode. For most participants, Manual Approve mode struck the right balance between giving Wearers control and giving Friends "live" in-the-moment access to their experience. Our concept of a shared camera is generalizable, and we believe that it can make in-the-moment experience sharing possible on any form of wearable camera interfaces.